# The Pioneer Anomaly:
# an inconvenient reality or NASA's 12 year misconception?

Paul ten Boom [1]

[1] *School of Physics, University of New South Wales, Sydney, NSW, 2052, Australia*

**Summary:** This paper discusses the likelihood of whether the Pioneer anomaly is due to 'mundane' systematic errors/effects or indicative of new or unappreciated physics. The main aim of this paper is to argue that recent publications suggesting that the anomaly is previously overlooked thermal recoil forces, which is in stark contrast to the earlier consensus (1998-2010), are open to questioning. Both direct and circumstantial evidence are examined, and the uncertainty or inaccuracy associated with observations of such a small magnitude effect is recognised. Whilst a non-systematic based anomaly appears to be very unlikely, by way of the awkwardness of the observational characteristics that would need to be modelled, the existence of other peripheral anomalous phenomena makes an outright dismissal of the anomaly unwise. Issues from the philosophy of science (and physics) are also tabled. In the interests of having a contingency plan, should future experiments provide support for a Pioneer-like anomaly, the type of unappreciated physics that could conceivably satisfy the awkward observational evidence is alluded to, albeit in a non-rigorous and cursory manner.

**Keywords:** Pioneer anomaly; spacecraft navigation; gravitation; Earth flyby anomaly; fundamental physics

## Introduction

The Pioneer 10 and 11 spacecraft, launched in 1972 and 1973 respectively, represent an ideal system to perform precision celestial mechanics experiments [1]. The radiometric Doppler tracking data of these spacecraft (in the outer solar system and beyond) indicated the presence of a small, anomalous, frequency blue-shift (relative to expectations) of $\approx 6 \times 10^{-9}$ Hz/s. This unmodelled Doppler 'drift', which is applicable to both spacecraft, has been interpreted as an anomalous deceleration – as compared to a clock (or time) acceleration effect [2]. The average value of this anomalous (Pioneer) deceleration ($a_P$) is $(8.74 \pm 1.33) \times 10^{-10}$ m/s$^2$. Note that this value is inclusive of a $+0.90 \times 10^{-10}$ m/s$^2$ bias due to various systematics (see Table II in Ref. 1), particularly the radio beam reaction force and the negative/inward contribution of onboard heat reflected off the spacecraft. The anomalous deceleration is an *offset* relative to predictions, with these predictions involving great complexity. Its direction is imprecisely determined, equally favoured are: Sun-pointing, Earth-pointing, and spin-axis directed. A path or velocity vector direction has also been considered consistent with the observations [1].

The spin stabilisation of these spacecraft, as compared to the three-axis stabilisation of the Voyager (1 and 2) spacecraft, makes them navigationally superior to the Voyagers. The only other spacecraft to have travelled beyond 20 AU, where the effect of solar radiation pressure has decreased to less than $4 \times 10^{-10}$ m/s$^2$ [3], is the New Horizons mission which is en route to encounter Pluto in July 2015. Somewhat surprisingly, the Pioneer 10/11 based data represents the 'cutting edge' of navigational precision and accuracy in the solar system. A future space mission dedicated to improving upon this level of navigational precision and accuracy is needed, especially considering that the Pioneers were launched in 1972 and 1973 respectively (i.e. 40 years ago). Whilst solar plasma and atmospheric effects upon the radio signal cannot be reduced, other features such as understanding spacecraft thermal characteristics and monitoring of the spacecrafts' spin history can be improved upon [4]. The inclusion of time delay based ranging data (in addition to the Doppler data) would be greatly beneficial, particularly because it is independent of the Doppler data and responds differently to the effects of solar plasma upon the raw data measurements [2].

From the first publication in 1998 [2], the preferred stance towards the anomaly has been to presume it is a 'mundane' systematic effect, either onboard or external to the spacecraft, or a computational effect. The comprehensive (50 page) publication in 2002 by Anderson et al. [1] doused this expectation. In response to this 'definitive' publication two clearly distinguished points of view were activated: those seeking a systematic explanation and those who believe the observations imply a need for some type of "new physics". The debate continues today with the current (2012) consensus firmly favouring a thermal recoil force based explanation.

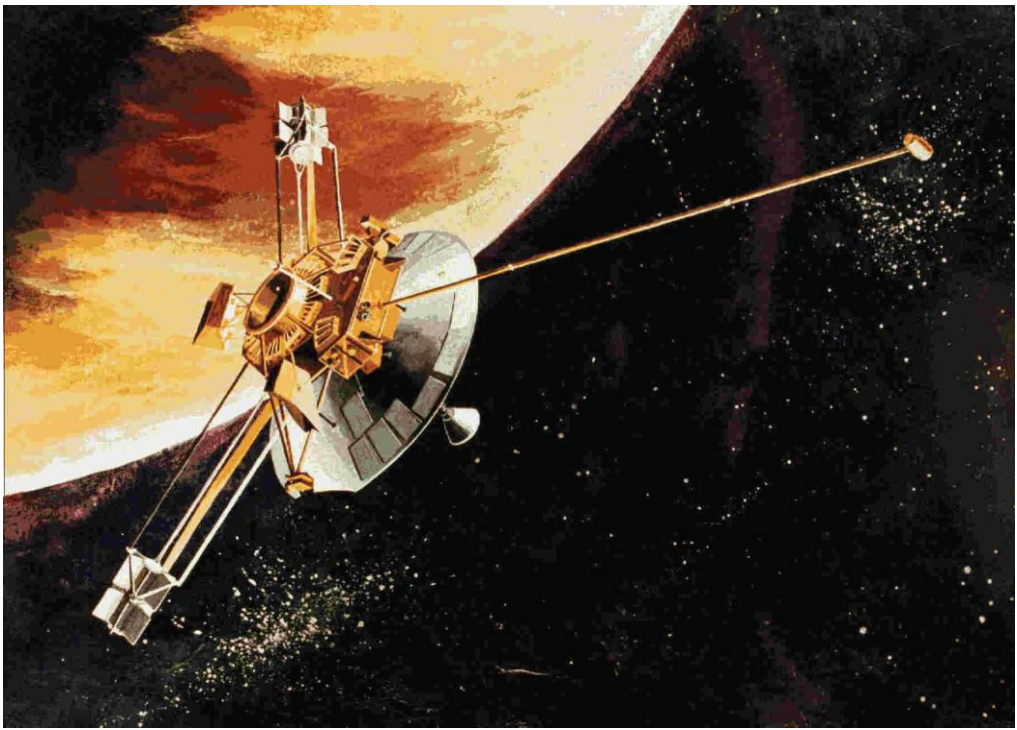

*Fig. 1: Artist's impression of the Pioneer 10/11 spacecraft (NASA Ames Research Center). Note the four SNAP-19 RTGs on two extended booms, as well as the magnetometer sensor.*

## Background information

The (outer solar system) Pioneer anomaly is potentially important because historically the motion of bodies in the solar system has provided fertile ground for scientific advancement.

Two fortuitous occurrences associated with the Pioneer spacecraft were: firstly (as regards navigational accuracy), the use of extended 3 metre long booms for the Radioisotope Thermoelectric Generators (RTGs), because of (ultimately unfounded) concerns involving radiation effects upon the onboard electrical equipment (see Fig. 1); and secondly, the success of Pioneer 10's Jupiter encounter allowed Pioneer 11 to be redirected across the solar system to Saturn (and beyond), such that it eventually travelled roughly in the opposite direction to Pioneer 10 (see Fig. 2). The New Horizons spacecraft has its RTGs much closer to the spacecraft, which does not bode well for its navigational accuracy. Accurate thermal modelling of the New Horizons spacecraft might possibly alleviate this issue, but it is highly unlikely that its navigational accuracy will approach that of the Pioneer 10 and 11 spacecraft.

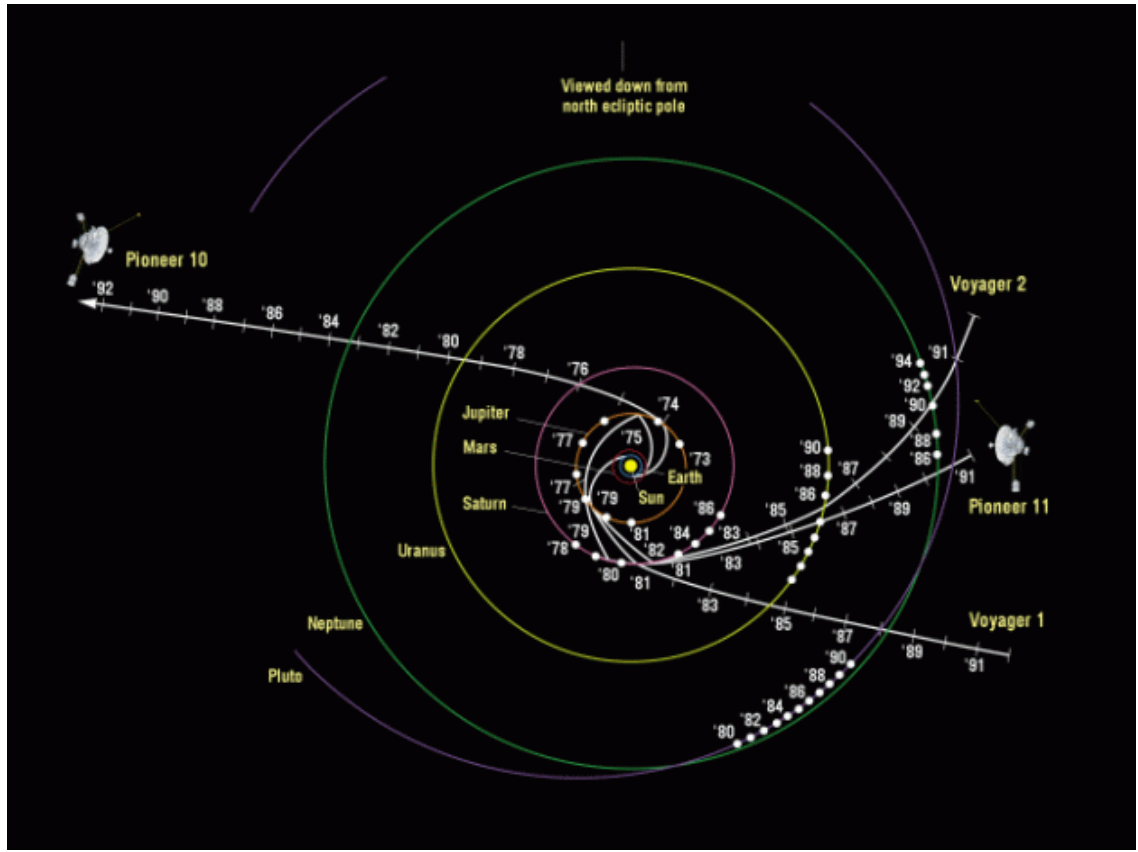

*Fig. 2: Hyperbolic trajectories of the Pioneer and Voyager spacecraft.*

Radiometric Doppler data acquisition and processing is a very complicated process, as is spacecraft 'orbit' determination. Feedback and long correlation times (of the order of 200 days) are involved [2][3].

The anomaly's magnitude is miniscule, but over the course of one million years a (path or) velocity vector based anomaly would experience a speed loss of greater than 27 km/s – which is roughly the orbital speed of the Earth around the Sun. If the anomaly is "real", in the sense of being a non-systematic and non-random error/effect, then the most awkward characteristic is that it cannot influence 'high' mass bodies such as: planets, moons, and large asteroids and comets [5][6][7]. Comets whose mass exceeds $10^{14}$ kg (approximately 7 km in diameter, assuming a comet density of $0.5 \times 10^3$ kg m$^{-3}$) are not affected [7], e.g. Halley's comet. If the anomaly is non-systematic based, it can be thought of as similar to the Poynting-Robertson effect [8] and the Yarkowsky effect [8], in the sense that it is an additional influence upon a body's motion that is not equally applicable to all sizes/scales of matter. Furthermore, it acts to perturb a body's motion in a minor manner – at least over cosmologically brief time scales. The crucial qualitative difference being that these effects are attributable to solar radiation or solar heating whereas the Pioneer anomaly (if 'real') is not.

# Evolution of the anomaly's interpretation between 1998 and 2012

**1998 to 2010: A plethora of hypotheses found wanting**

NASA was very cautious in not publishing until 18 years after the anomaly was first noticed in 1980, and four years after Michael M. Nieto (Los Alamos) and John D. Anderson (JPL, NASA) spoke in 1994 regarding the corroboration of deep space navigation with general relativity. Between 1998 and 2010, by way of a linear increase in the Doppler residual (cf. predictions/expectations), a (mean) *constant* anomaly was implied by the observations and analyses. This constancy was prominent in the first publication [1] and again in the definitive 2002 article [2], which featured an independent analysis from The Aerospace Corporation.

Between late 1998 and early 2010, all conceivable systematic based explanations, other than asymmetric thermal (radiation pressure) effects, were (able to be) ruled out. Furthermore, the awkward observational constraints ensured that all attempts at "new physics" were ultimately unsuccessful. Subsequently, the anomaly remained unresolved, unexplained and underdetermined.

The existence of a thermal (or heat) based explanation was argued for very early on [9], even though decay of the RTGs as well as reduction in the power supplied is inconsistent with a constant anomaly. The thermal recoil force hypothesis gained a firmer foothold in 2002 with Craig Markwardt's independent analysis [10] indicating that the timescale of his investigations could not preclude a jerk term (i.e. rate of change of acceleration). Note that he used a truncated data set (spanning 1987-1994) cf. the Anderson et al. 'extended' (highest quality) data set spanning 3 January 1987 to 22 July 1998 [1] – i.e. spanning roughly one solar cycle.

Interestingly, the original comprehensive analysis [1] had concluded that heat based effects (although present) were not significant. Indeed one French team's analysis of the 1987-1998 data set found the best fit to the data could only be a (mean) constant anomaly [11]. All other analyses (spanning 1998-2010) without exception confirmed the (mean) constancy of the anomaly, e.g. [10][12][13]. It was the noise/errors in the data that had permitted the postulation of a decaying anomaly, although an increasing anomaly could just as easily been entertained.

**Two very different schools of thought and the failure of "new physics"**

During these twelve years or so (1998-2010) the polarisation of the two points of view, and their interpretations of the data and peripheral evidence, became further entrenched. The anomaly's constancy was the thorn in the side of those who sought to explain the anomaly away, whereas the restriction to 'low' mass bodies thwarted the new physics contingent.

Today, the anomaly's interpretation is black and white, it is either heat based or it is not. The time for "I think the Pioneer anomaly might be due to …" is largely over. A non-heat based approach is unconventional, and (to date) its supporters have failed to deliver a fully viable and/or well-received hypothesis. Essentially, by way of default (and possibly impatience), the thermal recoil force hypothesis has evolved to become the strongly preferred stance/option. The earlier less rigorously examined data (see Fig. 7 in Ref. 1) suggested a mild decline in the anomaly between 1980 and 1986, in support of a heat systematic. Awkwardly, this earlier analysis also implied that Pioneer 11 en route to Saturn across the solar system (see Fig. 2) could have experienced a much smaller anomaly – about half the 'headline' magnitude.

**Factors weighing upon the anomaly's interpretation**

Over the years circumstantial 'evidence' such as: the anomaly's inconsistency with standard gravitational theory; the failure of new physics models; scepticism from galactic physicists and cosmologists; and the numerous successes of general relativity; has indirectly supported a (mundane) systematic cause of the anomaly, notwithstanding the observational evidence. It seems somewhat odd that nearly all galactic physicists, who have entertained the unfulfilled notion of dark matter for roughly 35 years, would not be more interested in a possible supplementation (rather than an extension) of standard gravitation. The extension of solar system based Newtonian Mechanics (with its dominant central sun) to galactic systems, with their approximately 100 billion stars, may not be a seamless extrapolation.

It is sensible to be conservative in science rather than outlandish, this is inarguable, but ultimately, hard won observational evidence should be a physicist's first priority. Compromising this point of view is the fact that the raw observations required significant processing. Simply citing general relativity's success fails to appreciate the uniqueness of the Pioneer experiment, i.e. a very small effect upon a body's motion over many *years*.

**2006 to 2012: The inclusion of older (post-1979 and pre-1987) data**

Between 2006 and 2012, with financial assistance from the Planetary Society, older pre-1987 data were retrieved and processed. This was a formidable task. Understandably, the initial expectation was that the extension of the data set would resolve the anomaly. Published results derived from this data [14][15] *suggest* a decaying anomaly (directed towards the Sun or Earth) in support of a heat based hypothesis. Concurring with this approach/agenda were two other teams: a Portuguese team [16] and a team at the Center of Applied Space technology and microgravity (ZARM) in Germany [17]. That a meager 65 watts or 3% asymmetry in total thermal radiation pressure is sufficient to attain the anomaly's magnitude [3] makes a thermal recoil force model very appealing. The main contributors are waste heat from (both) the RTGs and the electrical equipment inside the spacecraft's compartments.

Today, the Pioneer anomaly would appear to have become a non-issue were it not for the following mitigating circumstances. Firstly, John D. Anderson, the team leader of the original investigation [2], is not convinced. In a recent interview he argued that the new analysis has mis-modelled the solar radiation pressure [18]. By way of a process of elimination, John Anderson continues to pursue a non-heat or non-thermal recoil based approach [19][20]. Secondly, due to the greater number of maneouvres, and greater solar radiation pressure closer to the Sun, these new results (spanning 1980-1986) are necessarily of a lower quality than those comprising the 1987 to 1998 data set. Thirdly, the decades old data required translation into a common modern useable format. Not all of this data was recoverable. Spacecraft navigation is a complex process, requiring data feedback and long correlation times. The qualitative distinction of the new extended data set, as compared to the 'original' highest quality data set, is cause for concern.

The tiny magnitude of the Pioneer anomaly means that there is uncertainty in both the accuracy of the observations and the modelling thereof. Remarks within Turyshev, S.G. et al. [15] give lose support to this uncertainty: "... the addition of earlier data arcs, with greater occurrences of [spin axis orientation] maneuvers, did not help as much as desired (p.4) [15]." Further, "The gradually decreasing linear and exponential decay models [and the stochastic model] yield [only] *marginally* [italics added] improved fits when compared to the [steadily] constant acceleration model; … (pp. 3-4) [15]."

**Appraising and responding to the inclusion of the supplementary (1980-1986) data set**

It is not in the usual methodology of scientific enquiry to overturn a long established result by way of lesser quality older data, nor is it appropriate to disregard the opinion of a leading figure in the field. The appropriate response, in this author's opinion, would be to maintain both points of view and to deem the Pioneer Anomaly an "open issue". The new evidence is by no means decisive or conclusive, and a detailed account of the methods and results has (to date) not been published. On a lesser note, no direct physical (corroborative) testing has been done on the Pioneer H spacecraft, which was to be Pioneer 12 had NASA approved its out-of-the-ecliptic mission (1974 launch). It sits quietly in Washington D. C. at the National Air and Space Museum, albeit without its (radioisotope based) power supply.

The new/current consensus, i.e. the thermal recoil force explanation, is comforting in that it removes the theoretical conundrum that the original investigation raised. Of concern is the need for extensive parametrisation in establishing each of the three thermal models [14][16][17] – especially considering that the 'original' thermal modelling of the spacecraft was significantly different. Furthermore, the post-anomaly thermal modelling has the luxury of knowing the 'correct' (or sought) answer *before* setting the values of different parameters.

## Examining the case for a 'real' non-systematic based Pioneer anomaly

In the interests of comprehensiveness, and possibly scientific progress, it is worth examining the awkward theoretical constraints a non-systematic explanation would have to circumvent.

**Characteristics of a non-systematic based (or 'real') Pioneer anomaly**

One of the challenges confronting any non-systematic based model is the need to establish the characteristics that require modelling. A sound scientific methodology is to respect the hard won best quality observational evidence and then work backwards to a model from these observations, even though the peripheral theoretical circumstances are discouraging.

The primary characteristics of a Pioneer anomaly (1998-2010) that is not based upon systematic or random errors/effects are:
1. Only 'low mass' bodies are affected (i.e. $<10^{14}$ kg [7]). This implies a violation of the (weak) principle of equivalence if the anomaly is of a 'standard' gravitational nature.
2. Beyond 20 AU, where the effects of solar radiation are minimal, the mean value of the anomaly is constant.
3. There is temporal structure around the anomaly's mean constant value that exceeds the measurement error [21]. Furthermore, the 'annual' variation of the anomaly, i.e. the annual residual, has an amplitude that has proved difficult to account for [1][2], and a not quite annual period of approximately 355 days (or 0.0177 rad/day) [1][2][22][23].

The following (fourth) characteristic is more tenuous, it refines characteristics 2 and 3.
4. There is possibly a much lower value of the Pioneer anomaly for Pioneer 11 when it was between Jupiter and Saturn during its pre-1980 journey to Saturn across the solar system [1] (recall Fig. 2). The notion of a discontinuous onset of the Pioneer anomaly, activated around the time of Saturn encounter – possibly by way of a drag effect [24] – is unsupported by other aspects of the observational evidence.

**Response to these observational characteristics and theoretical constraints**

At first glance, the observational evidence would appear to defy any prospective model, including a conventional heat-based hypothesis. Some encouragement for the viability of a non-systematic based model can be found in the existence of the unresolved Earth flyby anomaly [25][26], and the empirical formula established regarding six of these Earth flybys [25]. Note that these flybys involve Earth based observations of Earth based gravitational assist events. Similar to the Pioneer anomaly there are two schools of thought on this issue, as there are regarding galactic flat rotation curves, with Occam's razor favouring conservatism.

Importantly, one should appreciate that (herein) General Relativity (GR) is not being considered wrong in any way, nor is GR in need of modification. This line of argument is inconsistent with the observational evidence/characteristics. A more appropriate response is to question the *scope* of general relativity, i.e. to examine whether all conceivable sources and means to non-Euclidean geometry or spacetime curvature are encompassed in the theory. The source type of the non-Euclidean geometry would need to be supplementary to and distinct from the standard sources of curved spacetime – as expressed in GR's stress-energy tensor. As regards the solar system, a process of elimination leaves a quantum mechanical (systemic) energy source as the only viable option. On first gloss, this implication sounds preposterous, but it may be in the interests of physicists to not disregard this unlikely and ambiguous option.

To entertain a supplementary form of non-Euclidean geometry, or 'gravitation' in the widest sense of the word, is tantamount to indirectly questioning the viability of the reductive agenda currently favoured by physicists. This agenda assumes a unification of the standard model's three forces and general relativity is the next big step in theoretical physics. Subsequently, physical reality may have two largely distinct domains/realms that interface and interact in a different manner. The relationship of the microscopic 'world' to the macroscopic world, especially regarding: quantum decoherence [27], quantum entanglement and non-locality, and kinematic based geometric phase offsets [28], remains open to interpretation. The external effect of analogue curved spacetime upon atomic or molecular quantum systems (in motion), particularly when it is below a minimum 'internal' energy level, would need reassessment.

It is not in the make-up of everyday physics that one needs to revisit the foundational aspects of gravitation, but in the case of a non-heat based Pioneer anomaly it seems unavoidable and appropriate. To "cast one's net" solely within accepted present-day physics, or by way of conventional 'modifications', will inevitably deny a non-systematic based Pioneer anomaly.

**Appreciating issues in the philosophy of science and the philosophy of physics**

The Pioneer anomaly could possibly be necessitating a deep rethink of fundamental concepts such as: space, time, mass, gravitation, and energy interactions. Discussions within the philosophy of science and philosophy of physics provide the deepest questioning of aspects of general relativity and its associated concepts. Although an enquiry of this type was signaled in the comprehensive 2002 analysis [1], from 1998 to 2012 there has been little or no philosophical based discussion in the literature pertaining to the Pioneer anomaly.

It is worth noting that on the road to a general theory of relativity, Einstein failed in his original objective of rendering acceleration relative (so as to 'satisfy' Mach's Principle). Related to this, both proper acceleration and rotation are notable in that they are inherently non-relative [29]. Furthermore, the physical (and ontological) status of the principle of general covariance and spacetime continue to be active topics of debate within the philosophy of science [30][31]. It is undeniable that spacetime is theoretically and (hence) observationally fundamental, but it may not be ontologically fundamental. The issues of (systemic) quantum entanglement and quantum non-locality are not necessarily independent of these concerns, in particular their apparent inconsistency with the special theory of relativity (SR) [32].

By way of these issues, it is conceivable that a solely local and relativistic approach to gravitation, in the sense of a metric based non-Euclidean geometry, is possibly an incomplete rendering of 'gravitation', in the widest possible sense of the world. One would need to be sympathetic to Vladimir Fock's preference for viewing SR as a "Theory of Invariance" and GR as a "Relativistic Theory of Gravitation" [33] in order to loosen the grip of GR, with its implicit assumption of reductionism to exclusively local causal effects. If physicists were able to define dark matter, characterise dark energy, and establish a 'progressive' (four force) unified field theory then this vague possibility of incompleteness would be totally inappropriate. Until such time, there is no harm done in exploring alternative options.

**The solar system's curious relationship to two unresolved issues in astrophysics**

The significance of fully understanding the motion of bodies in the solar system is undeniable. In the interests of thoroughness it appears prudent to mention two effects that (somewhat intriguing) may indicate that the solar system may be associated with additional field effects.

In both cases – noting that the later scenario is predominantly attributed to dark energy – the existence of a solar system based 'contaminating' foreground effect has been muted as a (low probability) possible explanation. Two distinct foreground effects should not be ruled out.

Firstly, there is the anomalous alignment of the cosmic microwave background (CMB) radiation anisotropy with respect to both the orientation of the solar system and its motion through space [34].

Secondly, the transition redshift from a decelerating expansion of the universe to an accelerating expansion of $z_t = 0.43 \pm 0.07$ [35] implies a lookback time (to transition) of $4.5 \pm 0.5$ billion years, for $H_0 = 71$ (km/s)/Mpc, $\Omega_M = 0.27$ and assuming a flat universe [see either Edward (Ned) Wright's (http://www.astro.ucla.edu/~wright/CosmoCalc.html) or Siobhan Morgan's (http://www.uni.edu/morgans/ajjar/) "Cosmic Calculators"]. This lookback time is remarkably consistent with the solar system's epoch of formation and early establishment, including when the largest moons of the solar system achieved spin-orbit resonance with respect to their host planets.

**Two qualifying remarks**

It is worth noting that the existence of the graviton particle and the extension of general relativity to the universe as a whole are actually provisional assumptions rather than assured reality. As such, observations implying a "flat" universe do not necessarily imply that dark matter and dark energy *must* exist. Their undeniably high likelihood relies on the assumption that physical reality and its theorisation, as currently understood, is largely finalised/complete.

The primary intention of this section's discussions has not been to outline a model for a non-systematic based Pioneer anomaly. It was merely to delineate aspects of the theoretical and observation 'landscape' surrounding the anomaly, particularly assumptions that might be obscuring alternative ideas. Resolute and final clarification of the Pioneer anomaly will conceivably occur years into the future. Until the issues discussed throughout this section are resolved it may be prudent to not entirely disregard the solar system based Pioneer anomaly.

# Conclusion

In support of John D. Anderson – team leader of the original Pioneer anomaly investigation – who is strongly sceptical of the thermal recoil force hypothesis [18], this paper has argued that a heat-based explanation does not constitute a watertight 'solution'. Although the (new) heat-based explanation overcomes the under-determinism that ensued from the comprehensive 2002 analysis, and is more likely and more scientific than all (to date) new physics based explanations, it too has its deficiencies. The major deficiency is its reliance upon an inferior quality extension (spanning 1980-86) of the highest quality data set (1987-98) to overturn the consensus view (1998-2010) of a (mean) *constant* anomaly. Having examined evidence that is both central and peripheral to the Pioneer anomaly, it is concluded that the anomaly should remain an open issue. Subsequently, the anomaly is neither an inconvenient truth, nor NASA's 12 year misconception. The ability of a dedicated space mission to significantly improve upon the navigational accuracy of the (40 year old) Pioneer spacecraft, and hence confirm or deny the veracity of the anomaly, was promoted.